\documentclass[aps,superscriptaddress,notitlepage,nofootinbib,twocolumn]{revtex4}
\usepackage[T1]{fontenc}
\usepackage[utf8]{inputenc}
\usepackage{amsmath,amssymb,mathtools,microtype,cancel,xcolor}
\usepackage[hidelinks]{hyperref}
\DeclareMathOperator{\trace}{tr}

\makeatletter
\renewcommand{\p@subsection}{\thesection.}
\makeatother
\usepackage[normalem]{ulem}
\begin{document}
\title{Emergent Gravity from Topological Quantum Field Theory: \\
Stochastic Gradient Flow Perspective away from the Quantum Gravity Problem}
\author{Andrea Addazi}
    %\email{addazi@scu.edu.cn}
    %\affiliation{Center for Theoretical Physics, College of Physics Science and Technology, Sichuan University, 610065 Chengdu, China}
    \affiliation{Institute of Astronomy and Astrophysics, School of Mathematics and Physics, Anqing Normal University, Anqing 246133, China}
    \affiliation{Laboratori Nazionali di Frascati INFN, Frascati (Rome), Italy, EU}
\author{Salvatore Capozziello}
    %\email{capozziello@unina.it}
    \affiliation{Dipartimento di Fisica ``E.\ Pancini'', Universit\`a  di Napoli ``Federico II'', Via Cinthia 9, 80126 Napoli, Italy}
    \affiliation{Scuola Superiore Meridionale, Largo San Marcellino, 10, 80138, Napoli, Italy}
    \affiliation{INFN Sezione di Napoli, Complesso Universitario di Monte Sant'Angelo, Edificio 6, Via Cintia, 80126, Napoli, Italy}
    \affiliation{Research Center of Astrophysics and Cosmology, Khazar University, Baku, AZ1096, 41 Mehseti Street, Azerbaijan}
 
 \author{Jinglong~Liu}
%\email{jinglong\_liu@sjtu.edu.cn}
\affiliation{Tsung-Dao Lee Institute, 1 Lisuo Road, Shanghai, 201210, China}
\affiliation{School of Physics and Astronomy, Shanghai Jiao Tong University, 800 Dongchuan Road, Shanghai 200240, China}
\affiliation{Center for Field Theory and Particle Physics \& Department of Physics, Fudan University, 200433 Shanghai, China}
\affiliation{Center for Astronomy and Astrophysics, Fudan University, 200433 Shanghai, China}   
    
\author{Antonino Marcian\`o}
   \email{marciano@fudan.edu.cn}
    \affiliation{Center for Field Theory and Particle Physics \& Department of Physics, Fudan University, 200433 Shanghai, China}
    \affiliation{Center for Astronomy and Astrophysics, Fudan University, 200433 Shanghai, China}  
    \affiliation{Laboratori Nazionali di Frascati INFN, Frascati (Rome), Italy, EU}
    \affiliation{INFN sezione di Roma ``Tor Vergata'', 00133 Rome, Italy, EU}
    
\author{Giuseppe Meluccio}
   % \email{giuseppe.meluccio-ssm@unina.it}
    \affiliation{Scuola Superiore Meridionale, Largo San Marcellino, 10, 80138, Napoli, Italy}
    \affiliation{INFN Sezione di Napoli, Complesso Universitario di Monte Sant'Angelo, Edificio 6, Via Cintia, 80126, Napoli, Italy}
\date{\today}

\author{Xuan-Lin~Su}
%\email{19110190015@m.fudan.edu.cn}
\affiliation{Center for Field Theory and Particle Physics \& Department of Physics, Fudan University, 200433 Shanghai, China}
\affiliation{Center for Astronomy and Astrophysics, Fudan University, 200433 Shanghai, China}
%TC:endignore

\begin{abstract}
\noindent
We propose a scenario according to which the ultraviolet completion of General Relativity is realized through a stochastic gradient flow towards a topological BF theory. Specifically, we consider the stochastic gradient flow of a pre-geometric theory proposed by Wilczek. Its infrared limit exists, and corresponds to a fixed point where stochastic fluctuations vanish. Diffeomorphism symmetries are restored in this limit, where the theory is classical and expressed by the Einstein-Hilbert action. The infrared phase then corresponds to the classical theory of General Relativity, the quantization of which becomes meaningless. Away from the infrared limit, in the pre-geometric phase of the stochastic gradient flow, the relevant fields of the Wilczek theory undergo stochastic fluctuations. The theory can be quantized perturbatively, generating corrections to the classical Einstein-Hilbert action. The stochastic gradient flow also possesses an ultraviolet fixed point. The theory flows to a topological BF action, to which non-perturbative quantization methods can be applied. Two phase transitions occur along the thermal time dynamics, being marked by: i) the breakdown of the topological BF symmetries in the ultraviolet regime, which originates the pre-geometric phase described by the Wilczek theory; ii) the breakdown of the parental symmetries characterizing the Wilczek theory, from which General Relativity emerges. The problem of quantizing the Einstein-Hilbert action of gravity finally becomes redundant. 
\end{abstract}

\maketitle

\section{Introduction}
\noindent 
Although the quest for a quantum theory of gravity remains after more than a century an open problem in high-energy physics, several authors have recently explored novel pathways that revolve around alternative scenarios rather than quantizing gravity. Models of emergent gravity, in which the Einstein-Hilbert action only emerges at infrared energy scales \cite{macdowell:unified,Liu:2023pok,Paganini:2025}, or to geometrize quantum mechanics \cite{Kibble:1978tm,Penrose,Isidro:2009ku,Singh:2023}, encoding gravity in the collapse of the wave function, or to recover a very stochastic origin of quantum mechanics itself \cite{Nelson,Calogero:1997ng, Adler}, eventually due to chaos effects \cite{Berry} or the percolation of high-energy degrees of freedom in the infrared regime \cite{Wilson}, have been developed. We show in this letter that these three seemingly distinct directions are in fact intertwined, converging toward a unified framework. 

%A pre-geometric (emergent) theory of gravity, proposed by Wilczek in \cite{wilczek:gauge}, undergoes a stochastic gradient flow driven by noise --- originating from chaos or high-energy degrees of freedom --- that permeates the infrared. General Relativity (GR) emerges as an infrared fixed point, while the ultraviolet (UV) limit corresponds to a topological theory. The intermediate, out-of-equilibrium regime is governed by Wilczek's pre-geometric theory, which exhibits an extended local symmetry beyond that of GR. The phase transition from this pre-geometric phase to GR was described by Wilczek as the spontaneous symmetry breaking (SSB) of a fundamental spacetime gauge symmetry \cite{addazi:pre-geometry}, where a classical metric structure arises from a Yang-Mills-type gauge theory without an \emph{a priori} metric or tetrads. The relevant fields are a gauge field \( A_\mu^{AB} \) ($A,B=1,\dots 5$), for either \( SO(1,4) \) or \( SO(3,2) \), and a Higgs-like scalar multiplet \( \phi^A \). When \( \phi^A \) acquires a vacuum expectation value (v.e.v.), the symmetry breaks to \( SO(1,3) \), recovering Einstein-Cartan gravity with a cosmological constant. We argue that the stochastic gradient flow accounts for both the IR transition to GR, via SSB, and the UV flow to a topological BF theory. In this context, where GR emerges as an effective theory at low energies, the traditional goal of quantizing GR becomes redundant, drawing an analogy with similar conclusions reached in previous literature \cite{Jacobson:1995ab,Padmanabhan:2007en,Verlinde:2010hp,Verlinde:2016toy}.

Within our framework, the pre-geometric theory of gravity proposed by Wilczek \cite{wilczek:gauge} is assumed to evolve under a stochastic gradient flow driven by noise originating from chaotic or high-energy degrees of freedom that permeate the infrared. General Relativity (GR) then emerges as an infrared (IR) fixed point, while the ultraviolet (UV) limit corresponds to a topological field theory. The intermediate regime is governed by Wilczek's gauge framework, characterized by an enlarged local symmetry beyond that of GR.\\

Within the framework of \cite{wilczek:gauge}, the transition to GR occurs via spontaneous symmetry breaking (SSB) of a fundamental spacetime gauge symmetry \cite{addazi:pre-geometry}. A classical metric structure dynamically arises from a Yang--Mills-type theory defined without an \emph{a priori} metric or tetrads. The fundamental fields are a gauge connection $A_\mu^{AB}$ ($A,B=1,\dots,5$), valued in $SO(1,4)$ or $SO(3,2)$, and a Higgs-like multiplet $\phi^A$, in the fundamental representation of the group. When $\phi^A$ acquires a vacuum expectation value, the symmetry breaks down to the stabilizer of either $SO(1,4)$ or $SO(3,2)$, namely $SO(1,3)$, yielding Einstein--Cartan gravity with cosmological constant. Indices of the fundamental representation of $SO(1,3)$ are hence denoted with Latin letters, running within the range $a,b=1, \dots, 4$. \\

In our picture instead, starting from the theory of Wilczek, the stochastic flow simultaneously accounts for the IR transition to GR (achieved in \cite{wilczek:gauge} via SSB) and the UV flow toward a topological BF theory. In this sense, GR appears as a low-energy effective phase of a more fundamental structure, rendering its direct quantization conceptually redundant, in analogy with emergent gravity-scenarios such as those ones described in seminal papers by Jacobson~\cite{Jacobson:1995ab}, Padmanabhan~\cite{Padmanabhan:2007en} and Verlinde~\cite{Verlinde:2010hp,Verlinde:2016toy}, and other relevant studies \cite{x26,x27,x28,x29,x30,x31,x32,x33}.\\

A microscopic derivation of the stochastic term is left open. Two complementary interpretations can be identified. First, in analogy with Nicolai-map constructions \cite{Malcha:2023wvf,Kadoh:2025qbe,Nicolai}, the noise may encode residual quantum fluctuations generated by integrating out high-energy (possibly matter or supersymmetric) sectors. Second, the noise may be intrinsically gravitational. As shown in \cite{lulli:stochastic}, multiplicative noise yields an effective cosmological constant proportional to the square of the stochastic fluctuation. The same structure arises geometrically from the quadratic (non-abelian) contributions of the projective connection in the Palatini formulation of the Einstein--Hilbert action. Away from equilibrium, torsional and distortional modes fluctuate around Euler--Lagrange configurations, naturally correlating stochastic and geometric sectors.\\

Multiplicative noise is essential for this non-trivial coupling, while the It\^o prescription ensures a consistent Markovian gradient-flow structure. Alternative stochastic schemes would modify the drift term but not the qualitative IR/UV mechanism. An independent, non-perturbative validation is provided by the topological approach of \cite{Asselmeyer-Maluga:2024zry}, where gauge confinement and topology change reproduce the same flow structure.\\

The plan of the paper is the following. In Sec.~\ref{WilTheo} we present the Wilczek pre-geometric theory and its canonical formulation. In Sec.~\ref{SPGF} we show that the stochastic gradient flow provides an alternative mechanism to SSB, in order to obtain, in the IR limit, GR from the theory of Wilczek. In Sec.~\ref{BFco} we recall a few elements of BF topological theories. In Sec.~\ref{BFGR} we review the deformation of a BF theory that implements simplicity constraints and hence provides GR. In Sec.~\ref{BFWT} we propose a deformed BF formulation of Wilczek theory. In Sec.~\ref{SGF} we investigate the stochastic gradient flow of the pre-geometric theory of Wilczek, and show that it flows in the IR limit toward GR, while in the UV limit toward a topological quantum field theory. We finally spell some conclusions in Sec.~\ref{Conc}. 

\section{Wilczek theory and its canonical analysis} \label{WilTheo}
\noindent 
Following \cite{macdowell:unified}, Wilczek proposed in \cite{wilczek:gauge} a different pathway to recast gravity as a gauge theory with parental gauge symmetry \( SO(1,4) \) or \( SO(3,2) \), and considered a SSB mechanism rather than an explicit symmetry breaking down to \( SO(1,3) \). The Lagrangian of the theory involves a term that reproduces both the Einstein-Hilbert action and the cosmological constant term within the SSB limit, and a SSB potential term, respectively  
\begin{equation} \label{Willy}
\mathcal{L}_\text{W} = k_\text{W} \epsilon_{ABCDE} \epsilon^{\mu\nu\rho\sigma} F_{\mu\nu}^{AB} \nabla_\rho \phi^C \nabla_\sigma \phi^D \phi^E,  
\end{equation}
\[
\mathcal{L}_\text{SSB} = -k_\text{SSB} v^{-4} |J| (\phi^2 \mp v^2)^2,  
\]  
where $F_{\mu\nu}^{AB}=\partial_{[\mu} A^{AB}_{\nu]}+f^{AB}_{\ \ \ \ CD \ EF}\, A^{CD}_\mu A^{EF}_\nu$ is the field strength of the connection $A^{AB}_\mu$ --- the square brackets denoting anti-symmetrization of indices (with coefficients $\pm1$), and the symbols $f^{AB}_{\ \ \ \ CD \ EF}$ either $SO(1,4)$ or $SO(3,2)$ structure constants --- \( \nabla_\mu \) the covariant derivative, and \( J \) a pre-geometric volume form. The conjugate momenta are recovered to be 
\begin{eqnarray}
&&\!\!\!\!\! \!\!\!\!\! \Pi_E = 2 \epsilon_{ABCDE} \epsilon^{0ijk} \nabla_k \phi^C \phi^D \times \nonumber\\
&&\!\!\!\!\! \!\!\!\!\! \left[ k_\text{W} F_{ij}^{AB} - 2 \text{sgn}(J) k_\text{SSB} v^{-4} \nabla_i \phi^A \nabla_j \phi^B (\phi^2 \mp v^2)^2 \right],  \nonumber
\end{eqnarray}
\[
\Pi^i_{AB} = 2k_\text{W} \epsilon_{ABCDE} \epsilon^{0ijk} \nabla_j \phi^C \nabla_k \phi^D \phi^E, \quad \Pi^0_{AB} = 0\,,  
\]  
leading to the form of the Hamiltonian density \cite{ACMM2}
\[
\mathcal{H}_\text{W} = \Pi^i_{AB} (\partial_i A_0^{AB} - 2 A_{C[0}^A A_{i]}^{CB}) - \Pi_A A_{B0}^A \phi^B.  
\]  
GR is then recovered via SSB, considering that $\phi^A \to v \delta_5^A$,  $A_\mu^{ab} \to \omega_\mu^{ab}$ and $A_\mu^{a5} \to m e_\mu^a$, hence implying the reduction to the Einstein-Hilbert action plus cosmological constant term, i.e.
\[
\mathcal{L}_\text{W} \to \frac{M_P^2}{2} e e^\mu_a e^\nu_b R_{\mu\nu}^{ab} - M_P^2 \Lambda e,  
\]  
where \( M_P^2 = -8k_\text{W} v^3 m^2 \) and \( \Lambda = \pm 6m^2 \). Within the same SSB limit, the conjugate momenta reduce to  
\[
\Pi^i_{ab} = -\frac{M_P^2}{4} \epsilon_{abcd} \epsilon^{0ijk} e^c_j e^d_k, \qquad \Pi^0_{ab} = 0.  
\]  
It is remarkable that the Hamiltonian recovered via the SSB matches the ADM gravity in the time gauge \( n_0 = 1 \), fixing \( e_0^0 = -1/N \)~\cite{ACMM2}. This is also the choice commonly adopted in the literature \cite{thiemann:modern} in order to canonically quantize GR using the Ashtekar variables. At the level of  Hamiltonian analysis {\it  \`a la}  Dirac, the theory has primary constraints \( Z_A \), \( Z^i_{AB} \), \( Z^0_{AB} \approx 0 \) and a secondary constraint \( \dot{Z}^0_{AB} \approx 0 \), and its total Hamiltonian is  
\[
\mathcal{H} = \lambda^A Z_A + \lambda_i^{AB} Z^i_{AB} + \lambda_0^{AB} Z^0_{AB} + \tilde{\lambda}_0^{AB} \dot{Z}^0_{AB}.  
\]  
The counting of the degrees of freedom is achieved by considering that the theory involves 90 dynamical variables, 20 gauge choices, 10 first-class constraints, and 44 second-class constraints. The $90$ dynamical phase space variables correspond to: twice (considering also the conjugated momenta) the four degrees of freedom per gauge potential times the ten internal dimensions of the gauge group --- namely $A_\mu^{AB}$, labeled by anti-symmetrized indices $A,B=1,2, \dots 5$; and twice the degrees of freedom of the Higgs-like pentaplet. \\

For completeness, and following the detailed analysis of Ref.~\cite{ACMM2}, we briefly summarize the canonical structure. The configuration variables are the spatial components of the $SO(1,4)$ (or $SO(3,2)$) connection $A_i^{AB}$ (with $i=1,2,3$ and $A,B=0,\dots,4$), yielding $3 \times 10 = 30$ fields, together with their conjugate momenta $\Pi^{i}_{AB}$, for a total of 60 canonical phase-space variables. In addition, the Higgs-like multiplet $\phi^A$ ($5$ components) and its conjugate momenta $\Pi_A$ contribute 10 further canonical variables, while the temporal components $A_0^{AB}$ ($10$ fields) act as Lagrange multipliers enforcing primary constraints. Altogether, this gives $90$ dynamical phase-space variables.\\

The constraint structure consists of 10 first-class constraints associated with the parent gauge symmetry (removing twice as much degrees of freedom), and 44 second-class constraints arising from the primary momenta relations and the symmetry-breaking sector. Gauge fixing removes 20 additional degrees of freedom. It is then possible to recover that theory yields 6 degrees of freedom in the phase space, hence 3 physical degrees of freedom, corresponding to a massless graviton and massive scalar mode~\cite{ACMM2}.

\section{Stochastic pre-geometric flow}  \label{SPGF}
\noindent 
In a seminal paper by Parisi and Wu \cite{ParisiWu}, stochastic quantization was extended to quantum fields, along the lines drawn in \cite{Nelson67, Nelson66, Guerra73}. The relaxation dynamics of fields toward equilibrium configurations that fulfill the Euler-Lagrange equations was assumed to be driven by the Langevin equation, and parametrized by an auxiliary scale variable, often referred to as thermal or stochastic time. The drift term in the Langevin equation is given by the functional derivative of the action, i.e. the differential operator entering the Euler-Lagrange equations, acting on the space of field configurations. The stochastic noise term encodes the effective contribution of unresolved high-energy degrees of freedom, whose influence percolates into the low-energy regime. 
\\

Quantum expectation values are substituted by statistical averages with respect to gaussian-weighted copies of the stochastic noise. Alternatively, considering the Fokker-Planck equation associated to the Langevin equation, quantum expectation values can be implemented as statistical averages with respect to the probability distribution that solves the Fokker-Planck equation \cite{Gardiner}. The equilibrium configurations of the system --- corresponding to the saddle points of the action and hence to the vanishing of the drift term --- are recovered in the infrared limit of the stochastic evolution.\\

The `pre-geometric' (Wilczek) phase can be connected to the infrared emergence of gravity if we consider a stochastic gradient flow in the pre-geometric variables. Remarkably, the same flow predicts not only the existence of an infrared fixed point, but also of an ultraviolet fixed point, in which the theory becomes topological.  
The fate of the symmetries of General Relativity (GR) and  topological theory are indeed related to the dynamics of pre-geometric  stochastic flows, respectively towards the infrared and the ultraviolet fixed point. 
To address this relation, we consider the stochastic (Ricci) geometry flow \cite{lulli:stochastic}, from the perspective of which we can investigate the gauge-fixing of the ADM metric induced by the SSB. For this purpose, we assume the thermal time to be proportional to $n_0$, the normal to  hyper-surfaces in the ADM formalism. The stochastic geometry flow then implements the evolution of  one-parameter family of metrics in the space of  gauge-fixing parameters of the theory, namely $g_{\mu\nu}(x;n_0)$, according to  
\begin{eqnarray}\label{eq:ricci-flow(broken)}
    -i\frac{\partial}{\partial n_0}g_{\mu\nu}(x;n_0)=&&\!\!\!\!\!\!-2R_{\mu\nu}(x;n_0) \nonumber \\
    &&\!\!\!\!\!\!+2\Lambda g_{\mu\nu}(x;n_0) +\xi_{\mu \nu}(x;n_0),
\end{eqnarray}
with $\xi_{\mu \nu}(x;n_0)$ stochastic noise \cite{lulli:stochastic}, and where the presence of  imaginary unit is due to the Lorentzian signature of the spacetime manifold. The result of the SSB in the pre-geometric theory, which was studied by some of us in \cite{ACMM2}, can now be interpreted, in geometric terms, as the convergence of the stochastic geometry flow towards a configuration characterized by a specific  gauge-fixing of the metric, namely $g_{\mu\nu}(x;1)$. \\

Eq.~\eqref{eq:ricci-flow(broken)} contains a specific combination of the geometric fields $g_{\mu\nu}$, $R_{\mu\nu}$ and the stochastic noise. A natural {\it ansatz} is that the stochastic pre-geometry flow equation (in the pre-geometric phase) must contain the stochastic noise and analogous combinations of fields than in Eq.~\eqref{eq:ricci-flow(broken)}. The latter ones are $P_{\mu\nu}\equiv\eta_{AB}\nabla_\mu\phi^A\nabla_\nu\phi^B$ and $w^{\mu}_{A}\equiv \pm\epsilon_{ABCDE}\epsilon^{\mu\nu\rho\sigma}\nabla_\nu\phi^B\nabla_\rho\phi^C\nabla_\sigma\phi^D\phi^E$, which now involve the Higgs pre-geometric multiplets $\phi^A$ and their covariant derivatives (using the parental gauge connections) --- see e.g. Ref.~\cite{addazi:pre-geometry}. \\

By inspection of the expression $w^{\mu}_{A}$, we realize that its component $w_0^0$ must be proportional to $n_0$ after the SSB. For further details, see \cite{ACMM2}. For these reasons, it is useful to define the geometric quantity $\bar{n}_0\equiv n_0/\bar{n}$. Within the time gauge, this corresponds to $\bar{n}^{-1}=6\prescript{(3)}{}{e}$. We then recognize that $\bar{n}_0$ corresponds to the SSB of the pre-geometric quantity $\bar{w}_0^0\equiv w_0^0/(v^4m^3)$. Finally, according to the correspondence principle\footnote{The ``correspondence principle'' refers to the dictionary proposed in Eq.~(32) of Ref.~\cite{addazi:pre-geometry}. In this letter, we have briefly summarized the identifications of Ref.~\cite{addazi:pre-geometry}, showing the SSB of Eq.~\eqref{eq:ricci-flow(unbroken)} down to Eq.~\eqref{eq:ricci-flow(broken-normalised)}.} \cite{addazi:pre-geometry}, the stochastic pre-geometric flow can be written, using the variables specified by the action proposed by Wilczek, as
\begin{eqnarray}\label{eq:ricci-flow(unbroken)}
    -i\frac{\partial}{\partial\bar{w}_0^0}P_{\mu\nu}=&&\!\!\!\! \mp8v^2m^2J^{-1}\eta_{BC}w_A^\rho\nabla_\mu\phi^CF_{\rho\nu}^{AB} \nonumber \\
    &&\!\!\!\! \pm6m^2P_{\mu\nu} + \xi_{\mu \nu}.
\end{eqnarray}
The SSB of this equation, highly non-linear in the pre-geometric fields, reproduces the stochastic geometry (Ricci) flow, automatically selecting \cite{ACMM2} the time gauge $\bar{n}_0=\bar{n}^{-1}$, namely 
\begin{eqnarray}
\label{eq:ricci-flow(broken-normalised)}
     -i\frac{\partial}{\partial\bar{n}_0}g_{\mu\nu}\!\!\! &=& \!\!\!-2\eta_{bc}e^\rho_ae^c_\mu(R^{ab}_{\rho\nu}\mp2m^2e^a_{[\rho}e^b_{\nu]})\pm6m^2g_{\mu\nu} \!+\! \xi_{\mu \nu}\nonumber \\
&=&\!\!\!-2R_{\mu\nu}+2\Lambda g_{\mu\nu}+ \xi_{\mu \nu}\,.
\end{eqnarray}
Thus, the dynamics of the pre-geometric fields --- described by the stochastic pre-geometric flow in Eq.~\eqref{eq:ricci-flow(unbroken)} --- yields the stochastic geometry (Ricci) flow, Eq.~\eqref{eq:ricci-flow(broken-normalised)}. This is achieved in the space of the gauge-fixing parameters of the theory, after the SSB, with the time gauge $\bar{n}_0=\bar{n}^{-1}$ being automatically `selected' at equilibrium as a consequence. While the pre-geometric stochastic flow interpolates among the infrared and the ultraviolet fixed points, applying the SSB corresponds to instantiating the phase transition among the pre-geometric and the infrared phases. It thus corresponds to `sitting close to' the infrared fixed point, where the (stochastic) fluctuations above classical GR, i.e. the infrared fixed-point equilibrium configuration, are governed by the stochastic (Ricci) geometry flow. As a consequence, to inspect the stochastic gradient flow \cite{lulli:stochastic} at any scale, we shall inspect its pre-geometric version. For this purpose, it is convenient to recast the pre-geometric theory in terms of a deformed BF theory \cite{celada:bf}.

\section{Constrained BF theories} \label{BFco}
\noindent 
BF theories are topological field theories, the extensions of which encode also the Palatini first-order formulation of GR~\cite{celada:bf,baez:introduction,baez:topological,witten:topological,atiyah:topological,ashtekar:review,freidel:topological,Capozziello:2012eu}. On a four-dimensional Lorentzian manifold equipped with a principal $G$-bundle, the action of a BF theory (without the cosmological constant term) is
\begin{equation}\label{eq:BF}
    S_\textup{BF}=\int\trace(B\wedge F),
\end{equation}
where the trace is computed over the Lie algebra indices of $G$. The dynamical system comprises the connection form $A$ for $G$ --- with curvature $F=dA+A\wedge A$, where the structure constants are suppressed in the patterns of indices contractions --- and a two-form $B$ taking values in the adjoint representation of $G$. The equations of motion of the theory correspond to the Gauss constraint, which is the generator of the gauge symmetries of $G$, and the curvature constraint, which imposes the flatness of the topological configurations. These equations are respectively
\begin{equation*}
\mathcal{D}B^{AB}=0,\qquad F^{AB}=0,
\end{equation*}
where $\mathcal{D}$ denotes the covariant derivative with respect to the connection $A$. The action Eq.~\eqref{eq:BF} is invariant under diffeomorphisms and gauge transformations, but also under shifts of the $B$ field of the form $B^{AB}\rightarrow B^{AB}+\mathcal{D}V^{AB}$ for any one-form $V$. The main consequence of this symmetry is the absence of local degrees of freedom and, thus, the topological invariance of the theory.

\section{BF reformulation of GR} \label{BFGR}
\noindent 
The reformulation of GR in the language of topological BF theory requires an extension of the BF action. This enables to impose constraints that unfreeze the dynamical degrees of freedom of GR out of the topological BF theory. Starting from the local Lorentz group $G=SO(1,3)$ as the gauge group of the theory, the Einstein-Hilbert action can be expressed as
\begin{equation}
S_\textup{EH}=\frac{M_\textup{P}^2}{4}\int\epsilon_{abcd}e^c\wedge e^d\wedge F^{ab}.
\end{equation}
This action describes a BF theory like that of Eq.\ \eqref{eq:BF}, but with the extra requirement that there exist some tetrad fields $e^a$ such that the form of the $B$ field is given specifically by
\begin{equation}\label{eq:simplicity-constraint}
    B_{ab}^{(\textup{EH})}=\frac{M_\textup{P}^2}{4}\epsilon_{abcd}e^c\wedge e^d\,. 
\end{equation}
The condition \eqref{eq:simplicity-constraint} is referred to as the ``simplicity constraint'' and breaks the topological invariance of the BF theory. The simplicity constraint can be directly imposed as a constraint in the action, by means of a Lagrange multiplier with adequate symmetries \cite{thiemann:modern}. This yields the Plebanski action \cite{plebanski:separation,de-pietri:plebanski},
\begin{equation}\label{eq:plebanski}
    S_\textup{P}=\int(B_{ab}\wedge F^{ab}+\Phi_{abcd}B^{ab}\wedge B^{cd})\equiv S_\textup{BF}+S_c^{(\textup{EH})},
\end{equation}
which is an extended BF theory provided with a constraint $S_c^{(\textup{EH})}$  involving the Lagrange multiplier $\Phi_{abcd}=-\Phi_{bacd}=-\Phi_{abdc}=\Phi_{cdab}$.  The extremization of $S_\textup{P}$ with respect to $\Phi_{abcd}$ imposes the condition
\begin{equation*}
    B^{(ab}\wedge B^{cd)}=0\,,
\end{equation*}
where the round brackets denote symmetrization of each pair of anti-symmetrized indices. This equation has four different solutions, two of which reduce $S_\textup{BF}$ to $S_\textup{EH}$ \cite{thiemann:modern,freidel:topological}, while the other two enable to recover the topological Holst term --- for the Hamiltonian analysis of BF theories and the Plebanski theory, see Refs.~\cite{escalante:dirac} and \cite{buffenoir:hamiltonian} respectively.

\section{BF reformulation of Wilczek theory} \label{BFWT}
\noindent 
The Wilczek action can also be rephrased as a BF theory, with a condition on the $B$ field that is the pre-geometric generalisation of GR's simplicity constraint, namely 
\begin{equation}\label{eq:pre-geometric-simplicity-constraint}
    B_{AB}^{(\textup{W})}=-2k_\textup{W}\epsilon_{ABCDE}\nabla\phi^C\wedge\nabla\phi^D\phi^E.
\end{equation}
The Palatini formulation of GR, in the first-order formalism, is then recovered after the SSB, thanks to the fact that
$B_{AB}^{(\textup{W})}\xrightarrow{SSB}B_{ab}^{(\textup{W})}=B_{ab}^{(\textup{EH})}$. From the perspective of the BF formalism, the pre-geometric simplicity constraints \eqref{eq:pre-geometric-simplicity-constraint} `unfreeze' the three dynamical degrees of freedom of the Wilczek action and prevent the theory from being topological BF. The SSB enables in the IR limit the recovery of GR from the pre-geometric theory proposed by Wilczek. In particular, this mechanism provides a new degree of freedom other than those of the graviton \cite{ACMM2}, i.e. that of the scalar field $\phi^5\equiv\rho$. This additional (out-of-equilibrium, with respect to the IR limit) degree of freedom can be key to achieve a UV completion of the theory. To address this point, we consider the stochastic pre-geometric flow of Wilczek theory, using the extended BF formalism, and show how both the IR phase transition to GR and the ultraviolet phase transition to a topological theory can be recovered.\\

The action of the Plebanski formulation of Wilczek theory, $S_\textup{P-W}$, is the pre-geometric version of the gravitational Plebanski action $S_\textup{P}$, provided the choice of either $SO(1,4)$ or $SO(3,2)$ as the gauge group rather than $SO(1,3)$, and a modified expression for the constraint. Therefore, we may express $S_\textup{P-W}$ as \begin{eqnarray}\label{actio}
        S_\textup{P-W}&=&\int(B_{AB}\wedge F^{AB}+\epsilon_{ABCDE}B^{AB}\wedge B^{CD}\phi^E ) \nonumber \\
        &\equiv& S_\textup{BF}+S_c^{(\textup{W})}.
    \end{eqnarray}
The extremisation of $S_\textup{P-W}$ with respect to $\phi_{E}$ imposes the condition
\begin{equation} \label{simpli}
    B^{(AB}\wedge B^{CD)}=0,
\end{equation}
which again requires the two-form $B$ to be a bi-vector. This can be solved by considering that the only one-form available at the pre-geometric level, in the BF formalism, is either the $SO(1,4)$ or the $SO(3,2)$ covariant derivative of the Higgs-like pre-geometric multiplet $\phi^A$, i.e. $\nabla \phi^A$. We then obtain two solutions $B^{AB}_{\rm Holst} \propto \pm \nabla \phi^A \wedge \nabla \phi^B$ that reproduce, in the IR symmetry broken phase, the topological Holst term for gravity, and two metric-dynamical solutions that, in the IR symmetry broken phase, reproduce the Einstein-Hilbert action, namely $B^{AB}_{\rm EH} \propto \pm \  \epsilon^{AB}_{\ \ CDE} \nabla \phi^C \wedge \nabla \phi^D \phi^E$. The latter solutions provide the Wilczek term in Eq.~\eqref{Willy}.

\section{Stochastic gradient flow} \label{SGF}
\noindent 
The stochastic pre-geometric flow --- for details see Appendix \ref{ItoApp} and \ref{solodet}, in which the seminal analysis of stochastic geometry flows developed in \cite{Rumpf} has been extended to the BF formulation of the theory proposed by Wilczek --- can be easily recovered from the pre-geometric Plebanski action in Eq.~\eqref{actio}, 
\begin{eqnarray}
\!\!\!\!\!\! {\frac{\partial A}{\partial s}}^{AB} &=& -i  \mathcal{D} B^{AB} + \xi_g A^{AB}\,, \label{uno}  \\ 
\!\!\!\!\!\! {\frac{\partial B}{\partial s}}^{AB} &=& -i F^{AB} -i \epsilon^{AB}_{\ \ \ \ CDE}\, B^{CD} \phi^E + \xi_f B^{AB} \label{due} \,, \\
\!\!\!\!\!\! {\frac{\partial \phi}{\partial s}}^A &=& -i \epsilon^{A}_{\ B CDE} \, B^{BC}\wedge B^{DE} + \xi_s \phi^A \label{tre} \,,
\end{eqnarray}
having denoted with $s$ the thermal time, and introduced for each field an assumption of stochastic multiplicative scalar noise \cite{lulli:stochastic}, namely $\xi_{g}^{AB}= \xi_g \,A^{AB}$, $\xi_{f}^{AB}= \xi_f\, B^{AB}$ and $\xi_{s}^{A}=\xi_{s} \,\phi^A$. The physical implications of the choice of multiplicative scalar noise were studied in \cite{lulli:stochastic,Lulli:2023fcl,Asselmeyer-Maluga:2024zry}. On the other hand, the origin of stochastic noise has been addressed from different perspectives. Summarizing among different studies, it can be: either related to chaos, e.g. involving the Anosov flow on principal bundles \cite{Asselmeyer_book}, well suited to be applied to the pre-geometric theory proposed by Wilczek; or connected, from the Wilsonian RG flow point of view, to super-symmetric \cite{Malcha:2023wvf}, or in general high-energy, degrees of freedom \cite{Kadoh:2025qbe} --- e.g. by means of the path integral approach and the use of the Nicolai map \cite{Nicolai}.\\

The stochastic pre-geometric flow of the Higgs-like scalar multiplet in Eq.~\eqref{tre} must be inspected for all values of the energy scale $s$, as different physical considerations apply for different energy regimes. The derivative of $\phi^A$ with respect to the thermal time is required to be zero when the theory becomes `classical' at `equilibrium' in the thermal time, as all the symmetries of the classical theory are recovered exactly. In our case, the `classical theory' can refer either to GR or the Wilczek theory. As for the stochastic noise, its effects are expected to be relevant only in the deep UV limit --- as the excitations of some underlying degrees of freedom of quantum gravity/chaotic nature --- as they drive the out-of-equilibrium dynamics of $\phi^A$. Therefore, $\xi_s$ is a dominant factor in the deep UV limit and a negligible factor otherwise. We can then distinguish four different energy regimes, with as many different theories being selected by the thermal-time evolution of $\phi^A$: deep UV, intermediate UV, low UV, and IR regimes. These correspond respectively to: topological BF theory, some deformed BF theory, Wilczek theory, and GR.\\

In the deep UV limit, the noise term is dominant and drags the orbits of the system away from the saddle point of the action. Thus Eq.~\eqref{tre} reduces to
\begin{equation} \label{BER}
\frac{\partial \phi^A}{\partial s} -\xi_s \phi^A\approx0\,.
\end{equation}
The solution of this equation --- see Appendix \ref{solodet} for the details of the derivation --- is found to be
\begin{equation}\label{solo}
\phi^A(s) \approx e^{-{\frac{\sigma^2_{\xi_s}}{2}} (s-s_0)} \phi_0^A(s),
\end{equation}
the limit $s\rightarrow\infty$ approaching asymptotically the UV fixed point, $s_0$ denoting the thermal time at the IR point, $\sigma^2_{\xi_s}$  the variance of $\xi_s$ and $\phi_0^A(s)$ a generic polynomial function of $s$. \\

The field configurations fluctuate stochastically around the classical saddle points, so that Eq.~\eqref{solo} must be regarded as describing only the deterministic drift component of the full dynamics. The complete stochastic evolution contains both drift and noise contributions, and their relative importance depends on the energy (or equivalently, on the stochastic time s). In the asymptotic UV regime ($s \to \infty$), stochastic fluctuations dominate over the drift terms. Since the drift terms reproduce the classical equations of motion --- and in particular enforce the simplicity constraints --- their dynamical suppression leads to an effective decoupling of the Higgs-like multiplet from the B-sector. In this limit, the constraint-enforcing sector becomes dynamically irrelevant and the theory flows toward a topological BF phase. The ``freezing'' of local degrees of freedom therefore occurs asymptotically: the higher the energy, the stronger the stochastic fluctuations (possibly of chaotic origin), and the more suppressed the interaction terms become. Conversely, as the energy decreases, stochastic fluctuations weaken and the interactions progressively ``unfreeze.''
\\

The intermediate UV regime corresponds to finite but large energies, $T_c \lesssim E \ll \infty$, or equivalently $\infty > s \gg M_{\mathrm P}^{-1}$. In this region, no term in Eq.~\eqref{tre} can be neglected and the system must be treated as fully coupled. The resulting dynamics describes a non-perturbative deformation of BF theory governed by stochastic gradient flow. Since this regime is intrinsically out of equilibrium, the simplicity constraint cannot be imposed sharply through a Lagrange multiplier; rather, it emerges dynamically only when the drift term regains dominance.\\

At lower (but still supercritical) energies, stochastic contributions become subleading. In particular, for energies slightly above the critical scale $T_c \lesssim M_{\mathrm P}$, thermalization in stochastic time sets in, implying $\partial_s \phi^A = 0$. The flow equation then reduces to the pre-geometric simplicity constraint \eqref{eq:pre-geometric-simplicity-constraint}, and the Wilczek phase is dynamically recovered.\\

Finally, below the critical scale $T_c$, the Higgs-like multiplet stabilizes both in stochastic time and in spacetime, approaching a constant configuration $\phi^A = v \delta^A_5$. In this regime, the pre-geometric simplicity constraint reduces to the geometric one, and General Relativity emerges in the IR. Importantly, this transition should not be interpreted as arising from a conventional spontaneous symmetry-breaking potential --- no such potential term is introduced in Eq.~\eqref{actio}. Rather, the ordering of the phases is induced dynamically by the stochastic (thermal-time) flow itself.
\\

This is akin to the case of QCD, in which the breaking of the chiral symmetry is not due to the Higgs field, but to the non-perturbative dynamics of the theory; the very same dimensional transmutation can be inspected through the lenses of the stochastic gradient flow of the interacting gluon and gravitational fields, the latter being responsible for topology changes in the geometric phase \cite{Asselmeyer-Maluga:2024zry}. In other words, in this context $v$ is not the vacuum expectation value of $\phi^A$ and $\phi^A$ is not really a Higgs-like field, because it does not implement any Higgs mechanism. Rather, it is the stochastic noise of the deformed BF theory that drives the evolution of $\phi^A$ and, as such, is responsible for `selecting' the theory that is valid at any energy regime. Gravity, meant as GR, only exists in the IR limit. The role of the scalar multiplet $\phi^A$, and thus the need for the Wilczek theory, is clear from this discussion: the topological trivialization of GR can only be achieved via the dynamics of this beyond-Standard-Model degree of freedom. Without this pre-geometric `clock' to interpolate between the various theories of this meta-theory, in fact, there would not be any (simple) way to link GR to a topological BF theory. 

\section{Conclusions} \label{Conc}
\noindent 
The pre-geometric theory proposed by Wilczek is endowed with three degrees of freedom, corresponding to the two modes of the graviton and one scalar mode. The scalar mode can be naturally associated to a conformal mode, which is geometrized in terms of a gradient flow. The spontaneous symmetry breaking of the parental symmetry in the pre-geometric theory provides, in the infrared limit, General Relativity, at energy scales in which quantization becomes meaningless. Within the same limit, the time gauge emerges naturally, and the formalism reproduces the classical phase space considered in the loop quantization schemes. The stochastic version of the gradient flow instantiates the infrared phase transitions towards General Relativity, corresponding to a fixed point of the stochastic flow. The ultraviolet fixed point of the pre-geometric theory also exists, and corresponds to a phase in which the theory becomes a topological BF. The Higgs-like multiplet scalar field interconnects between two phase transitions, the one happening from the ultraviolet topological sector of the theory to the pre-geometric regime, and the one from the pre-geometric regime to the infrared classical regime, proper of General Relativity. This scenario then provides a trivial ultraviolet fixed point in the topological framework, the quantization of which can be accomplished with standard non-perturbative methods.\\

A complementary route to the emergence of General Relativity is provided by the asymptotic safety program (see e.g. \cite{Eichhorn:2018yfc}), where gravity is assumed to admit a non-Gaussian ultraviolet fixed point under Wilsonian RG flow, with General Relativity recovered as the infrared effective theory. In contrast, in our framework, based on \cite{lulli:stochastic}, gravity is not quantized at the fundamental level: it emerges classically at the infrared fixed point of a stochastic gradient flow acting on a pre-geometric gauge theory, after spontaneous symmetry breaking of the Wilczek phase, while the ultraviolet completion is realized by a topological BF fixed point. Both approaches share the structural role of fixed points and universality, but differ in interpretation: asymptotic safety treats the metric as fundamental and quantum at all scales, whereas in our construction geometry becomes meaningful only after the symmetry-breaking transition. A possible bridge would be to compare universal data (critical exponents, scaling directions, matter back-reaction) and investigate whether the stochastic flow parameter admits an approximate mapping to a Wilsonian RG scale, as suggested by the analysis that some of use developed in \cite{Asselmeyer-Maluga:2024zry}.

%{\bf Acknowledgements}.
%\vspace{0.5cm}
%TC:ignore
\begin{acknowledgments}
\noindent 
AM wishes to acknowledge precious feedback from T. Asselmeyer-Maluga, J.M.~Isidro, C.F.~Paganini and T.P.~Singh. 
AA acknowledges the support of the National Science Foundation of China (NSFC) through the grant No.\ 12350410358; the Talent Scientific Research Program of College of Physics, Sichuan University, Grant No.\ 1082204112427; the Fostering Program in Disciplines Possessing Novel Features for Natural Science of Sichuan University, Grant No.\ 2020SCUNL209 and 1000 Talent program of Sichuan province 2021. SC and GM acknowledge the support of Istituto Nazionale di Fisica Nucleare, Sez.\ di Napoli, Iniziative Specifiche QGSKY and MoonLight-2. SC thanks the Gruppo Nazionale di Fisica Matematica (GNFM) of Istituto Nazionale di Alta Matematica (INDAM) for the support. AM acknowledges the support by the NSFC, through the grant No.\ 11875113, the Shanghai Municipality, through the grant No.\ KBH1512299, and by Fudan University, through the grant No.\ JJH1512105. This paper is based upon work from COST Action CA21136 -- Addressing observational tensions in cosmology with systematics and fundamental physics (CosmoVerse), supported by COST (European Cooperation in Science and Technology).
\end{acknowledgments}

\appendix 
\section{Choice of theory, uniqueness, and symmetry breaking}

\noindent 
Our starting point is the ``pre-geometric'' gauge theory proposed by Wilczek, which admits a fully gauge-invariant parent theory with gauge group $SO(1,4)$ or $SO(3,2)$, coupled to a Higgs-like multiplet $\phi^A$. Gravity (Einstein-Cartan with $\Lambda$) then arises by dynamical spontaneous symmetry breaking (SSB) down to $SO(1,3)$. Conversely, the original MacDowell-Mansouri (MM) formulation effectively implements the Lorentz projection at the level of the action, i.e. it breaks the parent symmetry explicitly rather than dynamically. This is precisely why Wilczek's framework is better suited for our stochastic-flow/emergence logic, where a symmetric UV completion and controlled phase transitions along the flow are central. Rewriting MM in BF-like variables (or attempting a BF reformulation) does not by itself cure the key conceptual issue: the need for a genuinely $SO(1,4)/SO(3,2)$-invariant parent theory with dynamical SSB, rather than an explicit symmetry reduction built into the action.\\

The ``Wilczek term'' in Eq.~\eqref{Willy} is unique once one imposes (i) invariance under the chosen parent group $G=SO(1,4)$ or $SO(3,2)$, (ii) a Higgs multiplet $\phi^A$ triggering SSB to $SO(1,3)$, and crucially (iii) linearity in the curvature 2-form (Palatini-like, mirroring the Einstein--Hilbert structure in first order form). Under these requirements the admissible invariant densities are strongly constrained, which is why the construction is ``minimal'' in the sense emphasized in the letter. Similarly, MM-type actions are ``unique'' only after adopting a different criterion, namely quadratic dependence in the curvature, which naturally produces the Einstein-Hilbert term plus additional topological contributions (e.g. Gauss-Bonnet-type terms) once the Lorentz projection is enforced.\\

The groups $SO(1,4)$ (de Sitter) and $SO(3,2)$ (anti-de Sitter) are the minimal simple enlargements of the Lorentz group $SO(1,3)$ that allow a symmetry-breaking pattern $G \to H \equiv SO(1,3)$ with a clear geometric interpretation (cosmological constant sign and dS/AdS structure). This minimality is exactly what we exploit: it reduces arbitrariness while still allowing a nontrivial parent symmetry whose breakdown yields Lorentz gravity in the IR. At the same time, choosing between $SO(1,4)$ and $SO(3,2)$ is physically meaningful (dS vs AdS), so ``uniqueness'' must be understood given the sign choice and the minimal parent group.\\

In both Wilczek-type and MM-type constructions the relevant invariant density is built using the Levi-Civita tensor of the fundamental representation of the parent group. Since $SO(1,4)$ and $SO(3,2)$ act naturally on a 5D internal space, the corresponding invariant $\varepsilon_{ABCDE}$ is 5D. After SSB to $SO(1,3)$, the usual 4D $\varepsilon_{abcd}$ emerges together with the vierbein sector, reproducing the standard first-order gravitational form.\\

Other breaking chains can be considered (e.g. larger groups with intermediate stages, or different Higgs representations), and in principle may yield extensions of GR (additional fields, torsion sectors, scalar-tensor-like structures, etc.). However, the purpose of the present work is to keep the framework minimal and tightly linked to the stochastic-gradient-flow picture: a minimal parent gauge symmetry $SO(1,4)$ or $SO(3,2)$, a minimal Higgs-like multiplet $\phi^A$, and the SSB $G\to SO(1,3)$ that produces Einstein-Cartan/GR as the IR phase, as we described in the previous sections.

\section{The It\^o calculus} \label{ItoApp}
\noindent 
The stochastic calculus provides the framework for analyzing systems influenced by randomness. At its core is the It\^o's Lemma, which generalizes the classical chain rule to stochastic processes\footnote{\textcolor{red}{About the physical consequences of the choice of the It\^o calculus, we direct interested  reader to \cite{Rumpf}.}}. Let $X_s$ follow the stochastic differential equation
\begin{equation}
    dX_s = \mu(X_s, s) \, ds + \sigma(X_s, s) \, dW_s,
\end{equation}
where $W_s$ is the standard Brownian motion, $\mu(X_s, s)$ is the drift, and $\sigma(X_s, s)$ is the diffusion coefficient. The term $\mu(X_s, s) \, ds$ governs deterministic evolution, while $\sigma(X_s, s) \, dW_s$ accounts for stochastic fluctuations. %If $\sigma$ is constant, the noise is additive; if it depends on $X_s$, the noise is multiplicative.

Interpreting the Gaussian white noise $\xi(s)$ as the formal derivative of the Brownian motion, $\xi(s) ds = dW_s$, we can write
\begin{equation}
    \frac{d X_s}{d s} = \mu(X_s, s) + \sigma(X_s, s) \, \xi(s)\,.
\end{equation}
At a saddle point where $\mu = 0$, this simplifies to
\begin{equation}
\frac{d X_s}{d s} = \sigma(X_s, s) \, \xi(s)\,.
\end{equation}

Let $f(x, s)$ be differentiable once in $s$ and twice in $x$. Then, the It\^o's Lemma provides for $x = X_s$
\begin{align} \label{eq:itocal}
    & df(X_s, s) = \left(\frac{\partial f}{\partial s}(X_s, s) + \mu(X_s, s) \frac{\partial f}{\partial x}(X_s, s) \right.\\  
    & \left.+ \frac{1}{2} \sigma^2(X_s, s) \frac{\partial^2 f}{\partial x^2}(X_s, s) \right) \, ds + \sigma(X_s, s) \frac{\partial f}{\partial x}(X_s, s) dW_s.\notag
\end{align}
Compared to the classical chain rule, the It\^o's Lemma includes an extra second-order term $\frac{1}{2} \sigma^2 \frac{\partial^2 f}{\partial x^2}$, due to $(dW_s)^2 = ds$.

\section{Approximate solution of $\phi^A$}\label{solodet}
\noindent 
We provide here more details about the derivation of the approximate solution of $\phi^A$, in Eq.~(13), following Sec.~VII of Ref.~\cite{lulli:stochastic}.

We first recall that, to derive Eqs.~(9)--(11), the stochastic processes $X$ are provided by the Langevin equation with multiplicative noise $\xi$, where $X = A^{AB}, B^{AB}\, {\rm or}\  \phi^A$. This means, for any variable $X$, we have
\begin{align}
    \frac{\partial X}{\partial s} = - i\frac{\delta S[X]}{\delta X} + \xi_X X.
\end{align}
%For $X = \phi^A$, we find Eq.~(11). From Eq.~(12), we derive 
%\begin{align}
%    \frac{\partial^2\phi^A}{(\partial s)^2} = \xi_s \frac{\partial\phi^A}{\partial s} = \xi_s^2 \phi^A\,.
%\end{align}
%We can then write 
%\begin{align}
%    d\phi^A = \frac{\partial \phi^A}{\partial s} ds + \frac{1}{2}\frac{\partial^2\phi^A}{(\partial s)^2} ds^2 = \frac{1}{2}\phi^A ds + \phi^A dW_s\,, 
%\end{align}
%where we have used the Wiener process such that $dW_s \equiv \xi_sds$ and $[dW_s]^2 =ds$. Using the It\^o's lemma with $X_s = \phi^A$, $f = \ln\phi^A$, and solving at the saddle point for which $B^{(BC}\wedge B^{DE)}=0$, we find
%\begin{align}
%    d\ln\phi^A = - \frac{1}{2}%(\phi^A)^2\frac{1}{(\phi^A)^2}
%    ds + dW_s\,.
%\end{align}
%The solution for $\phi_A$ is then easily recovered to be 
%\begin{align} \label{swi}
%    \phi^A = e^{(- 1/2)(s-s_0)}\phi_0^A\,, \qquad  \phi_0^A = {\rm const.}\times e^{W_s}\,.
%\end{align}
%In general, if the variance of the stochastic noise $\xi_s$ is not unitary, then $\xi_s = \sigma_\xi (dW_s/ds)$. Correspondingly  Eq.~\eqref{swi} becomes 
%\begin{align}
%    \phi^A = e^{(- 1/2)\sigma^2_{\xi}(s-s_0)}\phi_0^A,\ \ \phi_0^A = {\rm const.}\times e^{W_s}\,.
%\end{align}
For $X = \phi^A$, and for the saddle point of the action, i.e., $\delta S/\delta\phi^A =0$, the Langevin equation is left with the stochastic part
\begin{equation}
    d\phi^A = \phi^A\xi_s ds.
\end{equation}
In the Wiener process, for a increment defined as $dW_s = \xi_s ds$, if the variance of the noise is non-unitary, i.e., $\sigma_\xi \neq 1$, the process scales as $dW_s = \sigma_\xi \, dW_s$. This results in the It\^o stochastic differential equation as
\begin{equation}
    d\phi^A = \sigma_\xi\phi^A dW_s.\label{eq:itosde}
\end{equation}
Applying the It\^o calculus and Eq.~(B4), for a process given by Eq.~(C3) and $f=\ln\phi^A$, we have 
\begin{equation} \label{Ct}
    d\ln\phi^A = -\frac{1}{2}\sigma_\xi^2 ds + \sigma_\xi dW_s.
\end{equation}
Integrating Eq.~(C4) on both sides from $s_0$ to $s$, we derive
\begin{equation}
    \phi^A(s)=\phi^A(s_0)\exp\left(-\frac{1}{2}\sigma_\xi^2(s-s_0) + \sigma_\xi W_s\right)\,.
\end{equation}

Since $W_s = \int_{s_0}^s \xi_{s'} ds'$, in the infrared limit we can always write it as an analytical function of $(s-s_0)$ multiplied by a step function $\theta(s-s_0)$. In fact, when the infrared fixed point is reached, the stochastic flow picks up one preferred direction of the internal space, hence realizing symmetry breaking.

Notice finally that, within the assumption of having $\xi_f$, $\xi_g$ and $\xi_s$ sharing the same value of the variance --- e.g., one may reasonably assume the origin of stochastic noises to be same, i.e. $\xi_f=\xi_g=\xi_s$ --- the time-scale of the stochastic gradient flow is recognized to be the inverse of the square root of the cosmological constant \cite{lulli:stochastic}.

%TC:endignore
\end{document}